\documentclass[11pt,a4paper]{article} 

\usepackage[margin=1in]{geometry}
\usepackage{setspace}
\usepackage{lmodern}
\usepackage[T1]{fontenc}
\usepackage{microtype}

\usepackage[authoryear,round]{natbib}
\bibpunct{(}{)}{;}{a}{,}{,}  
\let\cite\citep

\usepackage{hyperref}
\usepackage{titlesec}
\usepackage{enumitem}
\usepackage{array}
\usepackage{booktabs}
\usepackage{tabularx} 

\usepackage{tikz}
\usetikzlibrary{arrows.meta,positioning,fit,shapes.misc,calc}
\definecolor{midgray}{RGB}{90,90,90}
\definecolor{lightgray}{RGB}{230,230,230}
\tikzset{
	box/.style={
		draw=midgray, rounded corners, fill=white,
		thick, align=center, inner xsep=7pt, inner ysep=8pt
	},
	group/.style={
		draw=midgray, rounded corners, fill=lightgray,
		align=center, inner xsep=8pt, inner ysep=6pt
	},
	>={Stealth[length=3mm]},
}

\hypersetup{ colorlinks=true, linkcolor=blue, citecolor=blue, urlcolor=blue, pdftitle={Proportionate Cybersecurity for Micro-SMEs: A Governance Design Model under NIS2}, pdfauthor={Roberto Garrone}
}\titleformat{\section}{\large\bfseries}{\thesection}{0.6em}{}
\titleformat{\subsection}{\normalsize\bfseries}{\thesubsection}{0.6em}{} 

\title{\textbf{Proportionate Cybersecurity for Micro-SMEs: A Governance Design Model under NIS2}} 

\author{
	Roberto Garrone\thanks{%
		This article builds upon conceptual insights developed within the EU Squad 2025 initiative 
		(Framework Service Contract HADEA/2024/OP/0010). 
		The author participated as a voluntary contributor to the project component 
		\textit{“Small Businesses, Strong Defences.”} 
		The present paper is an independent academic elaboration and does not reproduce, represent, 
		or claim ownership of any official Squad 2025 deliverable or intellectual property owned by 
		EUN Partnership AISBL or the European Commission. 
		All views expressed are those of the author alone and do not necessarily reflect 
		the official position of the contracting authority or the European Union.} \\
	\small University of Milano-Bicocca\\
	\small \texttt{roberto.garrone@unimib.it} \\
	\small	ORCID: \href{https://orcid.org/0000-0002-4661-0954}{0000-0002-4661-0954}
}

\date{November 2025\\
	\small Prepared for submission to the \textit{Journal of Cyber Policy (Taylor \& Francis)}}

\begin{document}
\maketitle

\doublespacing 
\begin{abstract}
Micro and small enterprises (SMEs) remain structurally vulnerable to cyber threats while facing capacity constraints that make formal compliance burdensome. This article develops a governance design model for proportionate SME cybersecurity, grounded in an awareness-first logic and informed by the EU Squad 2025 experience. Using a qualitative policy-analysis and conceptual policy-design approach, we reconstruct a seven-dimension preventive architecture---awareness and visibility, human behaviour, access control, system hygiene, data protection, detection and response, and continuous review---and justify each dimension's contribution to proportionality and risk reduction. We then map the model's regulatory scope and limits against the NIS2 Directive, Commission Implementing Regulation (EU) 2024/2690, the Digital Operational Resilience Act (DORA), the Cyber Resilience Act (CRA), and the EU Action Plan on Cybersecurity for Hospitals, clarifying which obligations are supported and which require complementary governance (e.g.,  role accountability, incident timelines, statements of applicability, sector-specific testing and procurement). The analysis argues that raising awareness is the fastest, scalable lever to increase cyber-risk sensitivity in micro-SMEs and complements---rather than replaces---formal compliance. We conclude with policy implications for EU and national programmes seeking practical, proportionate pathways to SME cyber resilience under NIS2.
\end{abstract} 

 
\section{Introduction}
\label{sec:intro} 

\subsection*{Journal Alignment and Policy Focus}
This article analyses the policy rationale and governance design underlying a recent EU initiative for SME cybersecurity. It derives design principles for proportionate, awareness-first governance under NIS2 and related instruments and clarifies regulatory scope and limits. The study aligns with the \textit{Journal of Cyber Policy}'s editorial focus on cybersecurity governance, regulatory implications, and the design of proportionate policy instruments rather than descriptive reporting. 

\subsection*{Framing and Positioning}
In contrast to purely technical or compliance-focused approaches, this article reconstructs and justifies a \textbf{governance design model} grounded in proportionality and prevention. It examines how preventive frameworks can translate complex regulatory expectations (NIS2, DORA, CRA) into scalable practices for micro- and small enterprises (SMEs). Cybersecurity has evolved into a hybrid domain where complex technology and fragmented politics converge, requiring governance arrangements that reflect both technical rationales and political considerations \cite{cavelty_wenger_2019, vanetmueller2013}. 

\subsection*{Core Policy Claims}
\begin{itemize} \item \textbf{Problem:} Small firms face structural barriers to formal compliance yet carry systemic cyber exposure. \item \textbf{Gap:} EU guidance exists but lacks a proportionate, behaviour-led implementation path for micro-SMEs. \item \textbf{Contribution:} A seven-dimension governance design model (awareness-first) and a regulatory scope map aligning practical prevention with NIS2/DORA/CRA limits.
\end{itemize} 

\subsection*{Research Aim and Questions}

\noindent\textbf{Aim.} 
Derive design principles for proportionate SME cybersecurity governance from the Squad 2025 experience.

\vspace{0.5em}
\noindent\textbf{Research Questions.}
\begin{enumerate}[label=\textbf{RQ\arabic*.},leftmargin=5em]
	\item Which preventive design dimensions best operationalise NIS2 proportionality for micro-SMEs?
	\item How do these map to regulatory scope and what additional actions are needed for compliance?
	\item What are the policy implications for EU and national SME programmes?
\end{enumerate}

\subsection*{Context}
European micro- and small enterprises encounter structural barriers to effective cybersecurity adoption — including limited dedicated budgets, fragmented technical support networks and low internal awareness. For instance, ENISA \cite{enisa2021_sme} found that many SMEs lack both cyber-skills and budgetary commitment, and the Eurobarometer \cite{eurobarometer2022} reported only around one-fifth of SMEs had provided cyber-awareness training in the previous year.

ENISA underlines that SMEs are particularly exposed to cyber threats and require controls proportionate to their size and capacities (\cite{enisa2017_smes_data, enisa2021_sme, digitalsme2025, eurobarometer2022, enisa2021_sme, enisa2024threat}. Recent EU programmes---including ENISA's \emph{Cybersecurity for SMEs Toolkit} and the Squad 2025 initiative---aim to bridge this gap through lightweight, prevention-oriented frameworks. Following ISO/IEC 27005:2022 and conceptual policy-design principles \cite{iso27005, gregor_hevner_2013}, this article presents an independent academic analysis. For micro-SMEs, behaviour-centric prevention with light-weight governance---awareness, multi-factor authentication (MFA), regular patching, secure backups, and simple incident checklists---delivers the highest marginal reduction in exposure per euro invested and operationalises the NIS2 Directive's proportionality principle. 

\subsection{Regulatory Governance and Proportionality as Design Principles}
Cybersecurity regulation has progressively shifted from prescriptive control lists toward
\textit{principles-based} governance frameworks that prioritise adaptive capacity and contextual judgement
over rigid compliance.  In this view, proportionality operates both as a \textit{normative constraint}—
ensuring fairness and legitimacy in administrative burdens—and as an \textit{instrumental heuristic}
for calibrating control intensity to organisational capacity and exposure.  
As argued by \citet{black2008}, principles-based regulation enables regulators to steer behaviour through
broad governance norms such as accountability and transparency rather than exhaustive rulebooks.  
\citet{baldwin2012} and \citet{hood2001} further locate proportionality within a wider logic of
risk-based governance, in which regulatory design balances expected benefits, enforcement feasibility,
and institutional capability.  

This logic aligns with broader policy‐design perspectives emphasising that regulatory robustness depends on the interplay between process design and outcome resilience \citep{howlett2011, capano2018}. 
In this sense, proportionality functions not only as a fairness principle but as a design parameter within the architecture of governance instruments, ensuring that adaptive capacity and feasibility coevolve.

For micro- and small enterprises, this dual role of proportionality becomes critical: limited resources
and heterogeneous digital maturity make uniform compliance unrealistic.  
Adopting \citet{mukherjee2021}'s perspective on administrative capacity, proportionate cybersecurity thus
requires a governance architecture that evolves with the organisation’s ability to internalise risk
management practices.  
The seven-dimension model derived from the Squad 2025 experience operationalises this logic by sequencing
preventive controls along a maturity gradient—from cognitive and behavioural enablers to institutionalised
accountability mechanisms.  
Conceptually, the framework contributes to regulatory-governance theory by translating abstract
principles into staged, preventive design elements that can be embedded in SME policy programmes.

\section{Policy Context}
\label{sec:policy}
The NIS2 Directive \cite{nis2} requires appropriate and proportionate risk-management measures. However, the operational meaning of proportionality for micro-SMEs remains contested. ENISA's guidance and threat landscape for SMEs offer actionable preventive levers but stop short of formal compliance mappings \cite{enisa2022, enisa2024threat}. National CSIRTs and ENISA reports highlight that the proportionality principle must balance minimal control burden with measurable prevention impact \cite{europeancommission2023}. Forms of principles-based regulation in cybersecurity mirror broader governance debates \cite{black2008, baldwin2012}. Complementary instruments---the Commission Implementing Regulation (EU) 2024/2690 \cite{ec2690}, the Digital Operational Resilience Act (DORA) \cite{dora2022}, the Cyber Resilience Act (CRA) \cite{cra}, and the EU Hospital Cybersecurity Plan (2025)---shape sectoral expectations and accountability. \textit{} Policy should prioritise usable, proportionate early-stage practices that create measurable exposure reduction before heavier compliance regimes are imposed. 

Recent analyses confirm that small firms continue to face structural capability gaps in cybersecurity maturity and implementation, despite the increasing availability of EU and national guidance \cite{rawindaran2023, sukumar2023, uchendu2021}.

Public toolkits such as ENISA’s \textit{SME Cybersecurity Toolkit} \citep{enisa2023b} and national references like the UK~NCSC’s \textit{Small Business Guide} \citep{ncsc2023} illustrate proportionate prevention pathways, though their voluntary uptake remains limited across Member States.

\section{Policy Analysis and Design Approach}
\label{sec:method}
We apply a qualitative, document-based policy analysis and conceptual policy-design approach, integrating: (i) comparison of ENISA, ISO 27005, and NIS2 control families \cite{iso27005, enisa2021_sme, nis2}; (ii) conceptual reconstruction of the model's preventive dimensions; and (iii) mapping to regulatory scope. This conceptual policy-design reconstruction also aligns conceptually with ENISA's Cybersecurity Maturity for SMEs (C-SME) framework \cite{enisa2020}, which promotes self-assessment and staged progression. Behavioural constraints (e.g.,  security fatigue) and compliance intent are grounded in established empirical work \cite{stanton2016, bulgurcu2010}. 

Empirical findings from organisational research further show that the adoption of cybersecurity practices in small firms is strongly shaped by capability and perceived relevance, reinforcing the need for adaptive, design-based governance models \cite{hasani2023, rawindaran2023} that evolve with emerging defensive paradigms. At the technical level, the evolution of active cyber defense — from passive prevention to deception-based and adaptive mechanisms — demonstrates how governance design increasingly integrates learning feedbacks from operational environments \cite{zhang_thing2021}.

Sources are public EU/ENISA/ISO documents and the author's professional insight; proprietary Squad 2025 content is not reproduced. Policy analysis paired with conceptual policy-design reconstruction enables transparent proportionality judgments and avoids over-reliance on prescriptive checklists. Policy analysis paired with conceptual policy-design reconstruction enables transparent proportionality judgments and avoids over-reliance on prescriptive checklists.

\paragraph{Derivation of the Model.}
The seven preventive dimensions originated from practice rather than from document analysis alone. They were first elicited from the author's direct observation and participation within the EU Squad~2025 initiative, where preventive governance measures for micro-SMEs were co-developed and iteratively refined. These experiential insights revealed that awareness and visibility systematically emerged as the most effective entry point for resilience building—both as a behavioural catalyst and as a prerequisite for any subsequent technical or procedural control. This empirical primacy of awareness justified the model’s awareness-first logic and informed the sequencing of the remaining dimensions: human behaviour, access control, system hygiene, data protection, detection and response, and continuous review. In a second analytical stage, each dimension was systematically evaluated against the public regulatory corpus—NIS2~Directive (Arts.~21–23), ISO/IEC~27005:2022, and ENISA guidance (2020–2024)—to assess its regulatory correspondence, proportionality, and limitations. This two-stage process (experiential derivation followed by normative mapping) ensured that the model remained empirically grounded while fully aligned with the preventive intent and proportionality principle embedded in EU cybersecurity regulation.

The resulting architecture therefore reflects both the lived priorities identified in field experience and the structured correspondence verified through regulatory analysis.

\begin{figure}[htbp]
	\centering
	\resizebox{\linewidth}{!}{%
	\begin{tikzpicture}[node distance=6mm]
		\node[group, minimum width=5cm] (stage1) {\textbf{Stage 1 — Empirical Derivation}\\
			\small From practice (EU Squad 2025)};
		\node[box, below=7mm of stage1] (inputs1) {\textbf{Inputs}\\
			Field observations, SME constraints, iterative co-design};
		\node[box, below=6mm of inputs1] (process1) {\textbf{Synthesis}\\
			Awareness-first governance logic};
		\node[box, below=6mm of process1] (output1) {\textbf{Output}\\
			\textit{Seven preventive dimensions}\\
			\scriptsize Awareness | Behaviour | Access | Hygiene | Data | Detect/Respond | Review};
		
		\node[group, right=3.8cm of stage1, minimum width=5cm] (stage2) {\textbf{Stage 2 — Normative Evaluation}\\
			\small Against regulation (NIS2, ISO/IEC 27005, ENISA)};
		\node[box, below=7mm of stage2] (inputs2) {\textbf{Inputs}\\
			Legal texts, standards, ENISA guidance};
		\node[box, below=6mm of inputs2] (process2) {\textbf{Mapping \& Tests}\\
			Correspondence to Arts.~21–23;\\ limits \& complementary actions};
		\node[box, below=6mm of process2] (output2) {\textbf{Output}\\
			Regulatory scope \& limits table\\ (coverage + extra compliance steps)};
		
		\foreach \i/\j in {stage1/inputs1,inputs1/process1,process1/output1,stage2/inputs2,inputs2/process2,process2/output2}
		\draw[->, thick, midgray] (\i) -- (\j);
		
		\draw[->, thick, midgray] (output1.east) -- node[above, sloped, yshift=2pt]{\scriptsize model evaluation} (inputs2.west);
		
		\draw[->, dashed, thick, midgray] (output2.west) .. controls +(0,-1) and +(0,-1) .. (process1.east)
		node[midway, below, sloped, xshift=1pt]{\scriptsize feedback / refinement};
		
	\end{tikzpicture}
}
	\caption{Two-stage methodological process in which Stage~1 derives the preventive governance model from practice, and Stage~2 evaluates it against relevant regulatory frameworks (NIS2, ISO/IEC~27005, and ENISA guidance). Source: Author’s elaboration, 2025.}
	
	\label{fig:methodboundary_side}
\end{figure}
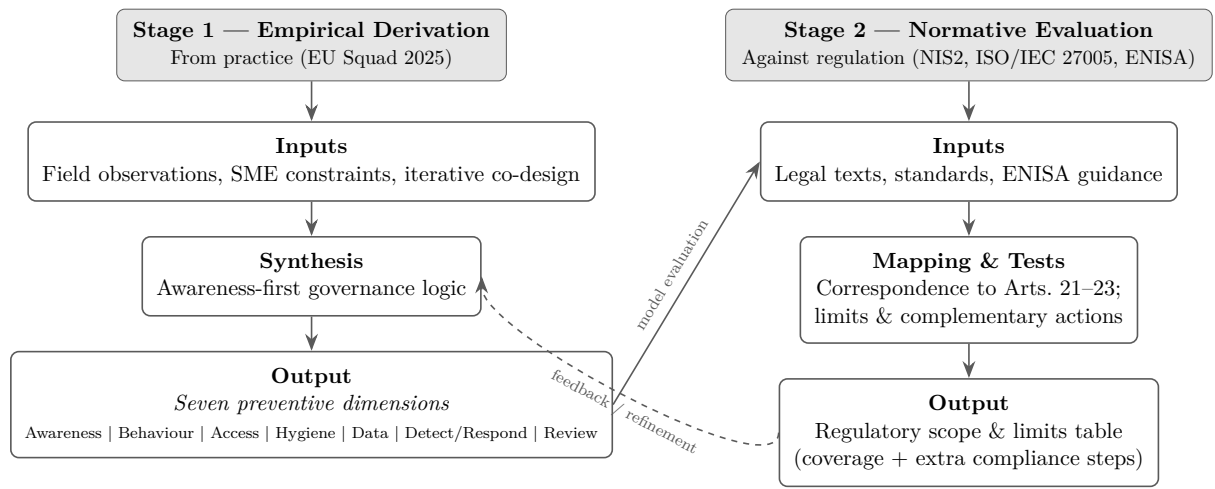

\section{Governance Design Model: Seven-Dimension Preventive Architecture}
\label{sec:model}
Building on the two-stage derivation process described above, this section presents the seven-dimension preventive architecture that emerged from the author’s experience and its subsequent regulatory evaluation.
Across SME-oriented guidance, seven preventive dimensions emerge as scalable governance components consistent with proportionality and prevention. 

Table~\ref{tab:framework} summarises the seven preventive dimensions that emerged from practical experience within the Squad~2025 initiative. Each dimension reflects an operational priority observed among micro-SMEs and has subsequently been evaluated for regulatory correspondence under the NIS2~Directive, ISO/IEC~27005, and ENISA guidance.

\begin{table}[h]
	\centering
	\renewcommand{\arraystretch}{1.15}
	\begin{tabularx}{\textwidth}{@{}p{0.9cm}p{3cm}X X@{}}
		\toprule
		\textbf{Step} & \textbf{Dimension} & \textbf{Preventive focus} & \textbf{Expected outcome} \\
		\midrule
		1 & Awareness and visibility & Asset/user mapping; simple risk identification & Exposure understanding; prioritised fixes \\
		2 & Human behaviour & Phishing hygiene, policy nudges, basic training & Fewer human-factor incidents; culture shift \\
		3 & Access control & MFA, least privilege, credential hygiene & Lower credential compromise risk \\
		4 & System hygiene & Updates, de-bloat, secure configurations & Smaller attack surface \\
		5 & Data protection & Encryption at rest/in transit; permissions & Controlled sensitive data access \\
		6 & Detection and response & Anomaly checks; escalation checklist & Faster containment and recovery \\
		7 & Continuous review & Self-assessment, drills, supplier checks & Ongoing maturity gains and accountability \\
		\bottomrule
	\end{tabularx}
	\caption{Seven-dimension governance design model for micro-SMEs, derived from field experience (Squad~2025) and evaluated against EU cybersecurity regulations (NIS2, ISO/IEC~27005, ENISA guidance).}
	\label{tab:framework}
\end{table}

The seven-dimension model complements ENISA's C-SME maturity approach \cite{enisa2020} and can support Member State toolkits. 

Behavioural and economic factors motivate this prioritisation: security fatigue and limited cognitive bandwidth constrain adoption (Stanton et al., 2016), while micro-SMEs optimise for maximum risk reduction per euro under capability and cost constraints \cite{anderson2006, mayadunne2016, gordonloeb2002, sukumar2023, enisa2021_sme}. The highest early-stage leverage comes from awareness, MFA, patching, backups, and checklists---a practical bridge from ENISA guidance to NIS2 objectives. 

\section{Policy Discussion and Implications}
\subsection{Policy Relevance}
The model operationalises proportionality by sequencing low-cost, high-impact controls before heavier obligations, aligning with NIS2's preventive intent \cite{nis2}. It provides a governance lens that national programmes can adopt to roll out awareness-first interventions at scale. Proportionate governance models let governments seed preventive capacity quickly while preserving paths to full compliance. 

\subsection{Regulatory Scope and Limits}
This section evaluates the correspondence between the preventive governance model and key EU regulatory instruments. While the model aligns conceptually with the preventive spirit of these frameworks, additional governance actions are necessary for complete legal conformity \cite{enisa2021_sme, ec2690, dora2022}. 


Table~\ref{tab:framework} presented the seven preventive dimensions derived from practice and structured as an awareness-first governance model. 
Table~\ref{tab:regscope} (below) extends this analysis by mapping each dimension against key EU regulatory instruments, clarifying the model’s coverage and the complementary governance actions required for full compliance.

\begin{table}[h]
	\centering
	\renewcommand{\arraystretch}{1.1}
	\begin{tabularx}{\textwidth}{@{}p{3.8cm}X X@{}}
		\toprule
		\textbf{Regulation} & \textbf{Model covers} & \textbf{Complementary actions for compliance} \\
		\midrule
		\textbf{NIS2 Directive (EU~2022/2555)} & Preventive awareness measures and baseline governance alignment (Arts.~21--23) & Appoint an accountable security officer; implement 24/72/30-hour incident reporting workflow; include supplier risk clauses in contracts. \\
		\textbf{Implementing Regulation (EU~2024/2690)} & Conceptual alignment with ISO/IEC control families & Draft a concise Statement of Applicability documenting control proportionality; define criteria for ``significant'' incidents. \\
		\textbf{Digital Operational Resilience Act (DORA, EU~2022/2554)} & Not sector-specific; general alignment with resilience objectives & Add ICT testing protocols, third-party criticality assessments, and detailed incident reporting if operating as a financial entity. \\
		\textbf{Cyber Resilience Act (CRA)} & User-side security hygiene and preventive awareness only & Incorporate CRA-aware procurement guidance when selecting or contracting hardware and software suppliers. \\
		\textbf{EU Hospital Cybersecurity Plan (2025)} & Awareness and preventive readiness components & Integrate coordination with ENISA’s sectoral Support Centre and national CSIRT-Health nodes to ensure healthcare-specific incident management. \\
		\bottomrule
	\end{tabularx}
	\caption{Regulatory correspondence of the seven preventive dimensions, showing coverage and complementary compliance actions required beyond the awareness-first governance model.}
	\label{tab:regscope}
\end{table}

As Table~\ref{tab:regscope} shows, the preventive architecture from Table~\ref{tab:framework} aligns conceptually with the core intent of NIS2 and related instruments but does not substitute for their formal obligations. 
This comparison highlights the proportional nature of the model: it translates the preventive spirit of EU cybersecurity law into operational practices that micro-SMEs can realistically adopt while maintaining a pathway toward full compliance.

Proportionality in cybersecurity governance acts both as a normative principle—ensuring fairness and preventing overregulation—and as a design heuristic that prioritises feasible prevention first. This dual nature situates proportionality at the intersection of regulatory legitimacy and institutional capacity, a tension central to principles-based regulation (Baldwin, Cave, \& Lodge, 2012). Proportionate governance echoes broader public-administration design literature on regulatory capacity and performance \cite{lodge2015}. At EU level, the NIS2 transposition guidance emphasises that proportionality requires context-specific discretion rather than uniform technical prescriptions \cite{europeancommission2023}. \textit{} Proportionate cybersecurity is attainable when prevention and governance evolve jointly and sectoral obligations are added where relevant. 

\subsection{Behavioural Foundations}
Behavioural economics and organisational research suggest that compliance intention increases with perceived fairness, efficacy, and cost realism \cite{bulgurcu2010}. Economic models show that SMEs' investment behaviour follows a cost--benefit logic sensitive to security externalities \cite{anderson2006, mayadunne2016}. Security adherence further depends on policy clarity and organisational reinforcement \cite{siponen2014}. Security fatigue limits the effectiveness of complex regimens \cite{stanton2016}. Therefore, simple, repeated behaviours---MFA, updates, backups---are most adoptable in micro-SMEs. ENISA's SME reports corroborate that guidance must be usable, concise, and staged \cite{enisa2021_sme, enisa2024threat}. Designs that minimise cognitive load and align with daily routines are more likely to persist and scale.

\subsection{Implications for EU Policy}
The model can inform national SME programmes, chambers of commerce initiatives, and sectoral support centres by providing a template for proportionate rollouts. It functions as a practical bridge between ENISA guidance and mandatory instruments, allowing policymakers to phase interventions from awareness to accountability \cite{cavelty_wenger_2019, vanetmueller2013}.

Consistent with this phased approach, national and EU programmes could integrate existing awareness toolkits from ENISA \citep{enisa2023b} and the NCSC \citep{ncsc2023} into proportionate rollout strategies. 
Such integration would operationalise the principles of policy robustness and design resilience \cite{capano2018}, ensuring continuity between preventive governance design and long‐term compliance trajectories.

 \textit{} Use the model as a scaffolding for tiered interventions: awareness now; accountability and incident governance as capacity grows. 

\section{Conclusion}
Proportionate cybersecurity for micro-SMEs is feasible when governance emphasises awareness-first prevention, behavioural realism, and staged compliance. The proposed model operationalises NIS2's preventive spirit and clarifies regulatory scope and limits, offering policymakers a practical path to scale SME resilience. 

\paragraph{Contribution and Limitations.}
Conceptually, this article contributes a governance-design framework that operationalises
the proportionality principle embedded in the NIS2 Directive for micro- and small enterprises.
By synthesising regulatory governance theory with insights from the Squad 2025 experience,
it translates abstract EU policy objectives into a practical, awareness-first architecture for
preventive cybersecurity. The model enriches the literature on principles-based regulation by
illustrating how proportionality can function simultaneously as a normative constraint and a
design heuristic for administrative capacity building. Its limitation lies in the absence of
empirical validation: while the framework has been conceptually accepted in an EU operational
setting, further research should assess adoption outcomes, cost-effectiveness, and cross-sector transferability through field studies and policy experiments.

Although grounded in EU regulation, the awareness-first proportionality logic parallels approaches such as the NIST Cybersecurity Framework tiers (U.S.) and the OECD Digital Security Guidelines. Cross-jurisdiction comparison could enhance policy coherence and inform global SME resilience strategies.

Future work should test adoption, cost-effectiveness, and risk reduction empirically. 

\section*{Acknowledgements}
The author contributed to the conceptual design of the seven-step preventive structure within the EU \emph{Squad 2025} initiative (HADEA Framework Service Contract 2024/OP/0010). All proprietary content of that deliverable remains \textcopyright{} EUN Partnership AISBL (2025). This article represents the author's independent academic interpretation and does not reproduce restricted material. 

The author acknowledges the use of \textit{ChatGPT (GPT-5, OpenAI, 2025)} to assist in language refinement, LaTeX formatting, and stylistic editing. 
The tool was not used for conceptual, analytical, or interpretive content generation. 
All content was reviewed and verified by the author to ensure accuracy and intellectual integrity.

\section*{Disclosure Statement}
No potential conflict of interest was reported by the author. 

\section*{Author Contribution}
Sole-authored paper; the author conceptualised, analysed, and wrote the manuscript. 

\section*{Data Availability}
\noindent Not applicable (conceptual analysis).

\bibliographystyle{apacite}

\begin{thebibliography}{40} 
	






\bibitem[Baldwin, Cave and Lodge(2011)]{baldwin2012}
Baldwin, R., Cave, M., \& Lodge, M. (2011).  
\newblock {\em Understanding Regulation: Theory, Strategy, and Practice} (2nd ed.).  
\newblock Oxford University Press.  
\newblock https://doi.org/10.1093/acprof:osobl/9780199576081.001.0001

\bibitem[Bulgurcu, Cavusoglu, and Benbasat(2010)]{bulgurcu2010}
Bulgurcu, B., Cavusoglu, H., \& Benbasat, I. (2010).
\newblock Information security compliance: An empirical study.
\newblock {\em MIS Quarterly, 34}(3), 523--548.
\newblock https://doi.org/10.2307/25750690

\bibitem[Cavelty and Wenger(2019)]{cavelty_wenger_2019}
Cavelty, M.~D., \& Wenger, A. (2019).
\newblock Cyber security meets security politics: Complex technology, fragmented politics, and networked science.
\newblock {\em Contemporary Security Policy, 41}(1), 5--32.
\newblock \url{https://doi.org/10.1080/13523260.2019.1678855}

\bibitem[Gregor \& Hevner(2013)]{gregor_hevner_2013}
Gregor, S., \& Hevner, A. R. (2013).
\newblock Positioning and presenting design science research for maximum impact.
\newblock {\em MIS Quarterly, 37}(2), 337--355.
\newblock https://doi.org/10.25300/MISQ/2013/37.2.01

\bibitem[Mukherjee et al.(2021)]{mukherjee2021}
Mukherjee, I., Howlett, M., \& Ramesh, M. (2021).
\newblock Policy capacity and effective policy design: A review.
\newblock {\em Policy Sciences, 54}(2), 345--366.
\newblock https://doi.org/10.1007/s11077-021-09420-8

\bibitem[Gordon \& Loeb(2002)]{gordonloeb2002}
Gordon, L. A., \& Loeb, M. P. (2002).
\newblock The economics of information security investment.
\newblock {\em ACM Transactions on Information and System Security, 5}(4), 438--457.
\newblock https://doi.org/10.1145/581271.581274

\bibitem[Mayadunne \& Park(2016)]{mayadunne2016}
Mayadunne, S., \& Park, S. (2016).
\newblock An economic model to evaluate information security investment of risk-taking small and medium enterprises.
\newblock {\em International Journal of Production Economics, 182}, 519--530.
\newblock https://doi.org/10.1016/j.ijpe.2016.09.018

\bibitem[Siponen, Mahmood, and Pahnila(2014)]{siponen2014}
Siponen, M., Mahmood, M.~A., \& Pahnila, S. (2014).
\newblock Employees' adherence to information security policies: A new perspective.
\newblock {\em Information \& Management, 51}(2), 217--224.
\newblock https://doi.org/10.1016/j.im.2013.08.006

\bibitem[van~Eeten and Mueller(2013)]{vanetmueller2013}
van~Eeten, M., \& Mueller, M. (2013).
\newblock Where is the governance in internet governance?
\newblock {\em New Media and Society, 15}(5), 720--736.
\newblock \url{https://doi.org/10.1177/1461444812462850}

\bibitem[Zhang \& Thing(2021)]{zhang_thing2021}
Zhang, L., \& Thing, V.~L.~L. (2021).
\newblock Three decades of deception techniques in active cyber defense – Retrospect and outlook.
\newblock {\em Computers \& Security, 106}, 102288.
\newblock https://doi.org/10.1016/j.cose.2021.102288

\bibitem[Anderson \& Moore(2006)]{anderson2006}
Anderson, R., \& Moore, T. (2006).
\newblock The economics of information security.
\newblock {\em Science, 314}(5799), 610--613.
\newblock https://doi.org/10.1126/science.1130992


\bibitem[Black(2008)]{black2008}
Black, J. (2008).
\newblock Forms and paradoxes of principles-based regulation.
\newblock {\em Capital Markets Law Journal, 3}(4), 425--457.
\newblock https://doi.org/10.1093/cmlj/kmn026

\bibitem[Stanton et al.(2016)]{stanton2016}
Stanton, B., Theofanos, M., Prettyman, S. S., \& Herley, C. (2016).
\newblock Security fatigue.
\newblock {\em IT Professional, 18}(5), 26--32.
\newblock https://doi.org/10.1109/MITP.2016.84

\bibitem[Hood, Rothstein, \& Baldwin(2001)]{hood2001}
Hood, C., Rothstein, H., \& Baldwin, R. (2001).
\textit{The Government of Risk: Understanding Risk Regulation Regimes}.
Oxford: Oxford University Press.
\newblock https://doi.org/10.1093/0199243638.001.0001

\bibitem[Capano \& Woo(2018)]{capano2018}
Capano, G., \& Woo, J.~J. (2018). 
\newblock Designing policy robustness: Outputs and processes.
\newblock {\em Policy and Society, 37}(4), 422--440.
\newblock https://doi.org/10.1080/14494035.2018.1504488

\bibitem[Rawindaran et al.(2023)]{rawindaran2023}
Rawindaran, N., Jayal, A., Prakash, E., \& Hewage, C. (2023).
\newblock Perspective of small and medium enterprise (SMEs) and their relationship with government in overcoming cybersecurity challenges and barriers in Wales.
\newblock {\em International Journal of Information Management Data Insights, 3}(2), 100191.
\newblock https://doi.org/10.1016/j.jjimei.2023.100191

\bibitem[Sukumar et al.(2023)]{sukumar2023}
Sukumar, A., Choudhary, A., Huatuco, L. H., Garza-Reyes, J. A., \& Kumar, V. (2023).
\newblock Cyber risk assessment in small and medium-sized enterprises: A multilevel decision-making approach.
\newblock {\em Risk Analysis, 43}(12), 2662--2682.
\newblock https://doi.org/10.1111/risa.14092

\bibitem[Uchendu et al.(2021)]{uchendu2021}
Uchendu, B., Nurse, J.~R.~C., Bada, M., \& Furnell, S. (2021).
\newblock Developing a cyber security culture: Current practices and future needs.
\newblock {\em Computers \& Security, 109}, 102387.
\newblock https://doi.org/10.1016/j.cose.2021.102387

\bibitem[Howlett(2011)]{howlett2011}
Howlett, M. (2011). 
\textit{Designing Public Policies: Principles and Instruments}. 
\newblock Routledge.
\newblock Retrieved from \url{https://www.taylorfrancis.com/books/mono/10.4324/9781315232003/designing-public-policies-michael-howlett}

\bibitem[Hasani et al.(2023)]{hasani2023}
Hasani, T., O’Reilly, N., Dehghantanha, A., Levallet, N., \& Rezania, D. (2023).
\newblock Evaluating the adoption of cybersecurity and its influence on organizational performance: A study of SMEs.
\newblock {\em SN Business \& Economics, 3}, 97.
\newblock https://doi.org/10.1007/s43546-023-00477-6

\bibitem[Lodge and Wegrich(2015)]{lodge2015}
Lodge, M., \& Wegrich, K. (2015).
\textit{The Problem-Solving Capacity of the Modern State: Governance Challenges and Administrative Capacities}.
Oxford University Press.
https://doi.org/10.1093/acprof:oso/9780198716365.001.0001




\bibitem[ISO/IEC(2022)]{iso27005}
ISO/IEC. (2022).
\newblock {\em Information security, cybersecurity and privacy protection — Guidance on managing information security risks (ISO/IEC 27005:2022)}.
\newblock International Organization for Standardization.
\newblock Retrieved from \url{https://www.iso.org/standard/80585.html}

\bibitem[CISA(2023)]{cisaEssentials2023}
CISA. (2023).
\textit{Cyber Essentials for Small Businesses}. U.S. DHS.

\bibitem[ENISA(2020)]{enisa2020}
ENISA. (2020).
\textit{Cybersecurity Maturity for Small and Medium Enterprises (C-SME) Framework}. European Union Agency for Cybersecurity. 

\bibitem[ENISA(2021)]{enisa2021}
ENISA. (2021).
\textit{Supporting the NIS Directive Implementation: Guidance for SMEs and Public Administrations}. European Union Agency for Cybersecurity. 

\bibitem[ENISA(2022)]{enisa2022}
ENISA. (2022).
\textit{Good Practices for SME Cybersecurity: National Initiatives Supporting SMEs}. European Union Agency for Cybersecurity. 

\bibitem[ENISA(2024)]{enisa2024threat}
ENISA. (2024).
\newblock {\em Threat Landscape for SMEs 2024}.
\newblock European Union Agency for Cybersecurity.
\newblock Retrieved from \url{https://www.enisa.europa.eu/publications/enisa-threat-landscape-2024}

\bibitem[European Commission(2022)]{nis2}
European Commission. (2022).
\textit{Directive (EU) 2022/2555 (NIS2)}. Official Journal of the European Union. 

\bibitem[European Commission(2024)]{ec2690}
European Commission. (2024).
\textit{Commission Implementing Regulation (EU) 2024/2690}. Official Journal of the European Union. 

\bibitem[European Commission(2023)]{europeancommission2023}
European Commission. (2023).
\textit{NIS2 Transposition Guidance: Implementation Support for Member States}. Directorate-General for Communications Networks, Content and Technology. 

\bibitem[European Parliament and Council(2022)]{dora2022}
European Parliament and Council. (2022).
\textit{Regulation (EU) 2022/2554 on Digital Operational Resilience for the Financial Sector (DORA)}. OJ L 333. 

\bibitem[European Commission(2022)]{cra}
European Commission. (2022).
\textit{Cyber Resilience Act}. COM/2022/454 final. 

\bibitem[ENISA(2023b)]{enisa2023b}
European Union Agency for Cybersecurity (ENISA). (2023b). 
\textit{SME Cybersecurity Toolkit}. 
Available at: \url{https://www.enisa.europa.eu/publications/sme-cybersecurity-toolkit}.

\bibitem[NCSC(2023)]{ncsc2023}
National Cyber Security Centre (NCSC). (2023). 
\textit{Small Business Guide: How to Improve Cyber Security}. 
UK Government. 
Available at: \url{https://www.ncsc.gov.uk/collection/small-business-guide}.

\bibitem[European~Commission(2022)]{eurobarometer2022}
European~Commission. (2022).
\newblock {\em Survey on the Experience of SMEs with Cybercrime (Flash Eurobarometer No.~500)}.
\newblock Directorate-General for Communications Networks, Content and Technology.
\newblock Retrieved from \url{https://europa.eu/eurobarometer/surveys/detail/2280} 

\bibitem[ENISA(2021)]{enisa2021_sme}
ENISA. (2021).
\newblock {\em Cybersecurity for SMEs: Challenges and Recommendations}.
\newblock European Union Agency for Cybersecurity.
\newblock Retrieved from \url{https://www.enisa.europa.eu/sites/default/files/publications/ENISA%20Report%20-%20Cybersecurity%20for%20SMES%20Challenges%20and%20Recommendations.pdf}

\bibitem[ENISA(2017)]{enisa2017_smes_data}
ENISA. (2017).
\newblock {\em Guidelines for SMEs on the Security of Personal Data Processing}.
\newblock European Union Agency for Cybersecurity.
\newblock Retrieved from \url{https://www.enisa.europa.eu/publications/guidelines-for-smes-on-the-security-of-personal-data-processing}

\bibitem[DigitalSME(2025)]{digitalsme2025}
Digital~SME~Alliance. (2025).
\newblock {\em Position Paper – The EU Cybersecurity Act and the Role of Standards for SMEs}.
\newblock European Digital SME Alliance.
\newblock Retrieved from \url{https://www.digitalsme.eu/cybersecurity-privacy/}


\end{thebibliography}

\end{document}